\documentclass[aps,prl,twocolumn,groupedaddress,amssymb,amsmath,showpacs]{revtex4}
\usepackage{dcolumn}
\usepackage{graphicx}

\begin{document}


\title{Spin Polarization and Transport of Surface States in the Topological Insulators Bi$\mbox{\boldmath $_2$}$Se$\mbox{\boldmath $_3$}$ and Bi$\mbox{\boldmath $_2$}$Te$\mbox{\boldmath $_3$}$ from First Principles}

\author{Oleg V. Yazyev}
\affiliation{Department of Physics, University of California, Berkeley, California 94720, USA}
\affiliation{Materials Sciences Division, Lawrence Berkeley National Laboratory, Berkeley, California 94720, USA}
\author{Joel E. Moore}
\affiliation{Department of Physics, University of California, Berkeley, California 94720, USA}
\affiliation{Materials Sciences Division, Lawrence Berkeley National Laboratory, Berkeley, California 94720, USA}
\author{Steven G. Louie}
\affiliation{Department of Physics, University of California, Berkeley, California 94720, USA}
\affiliation{Materials Sciences Division, Lawrence Berkeley National Laboratory, Berkeley, California 94720, USA}

\date{\today}


\begin{abstract}
We investigate the band dispersion and the spin texture of topologically protected surface 
states in the bulk topological insulators Bi$_2$Se$_3$ and Bi$_2$Te$_3$ by first-principles 
methods. Strong spin-orbit entanglement in these materials reduces the spin-polarization
of the surface states to $\sim$50\% in both cases; this reduction is absent in simple models 
but of important implications to essentially any spintronic application. We propose a way of 
controlling the magnitude of spin polarization associated with a charge current in thin films 
of topological insulators by means of an external electric field. The proposed dual-gate 
device configuration provides new possibilities for electrical control of spin.
\end{abstract}

\maketitle

The recently discovered three-dimensional topological insulators (TIs)~\cite{fukanemele,moorebalents,hsieh08}
realize an unconventional electronic phase driven by strong spin-orbit interaction (SOI). 
The striking feature of these heavy-element materials is the existence of metallic Dirac 
fermion surface states characterized by an intrinsic spin helicity: the wavevector of the 
electron determines its spin state.  A net spin density is thus produced upon 
driving a charge current at the surface of a TI.  Considerable effort has been devoted 
recently to possible applications of this property of TIs to novel electronic devices, 
e.g., to spintronics~\cite{garatefranz} and topological quantum computing~\cite{fukaneprox3d}.

Understanding the properties of bulk TIs has mostly relied on few-band phenomenological
models which treat the helical surface states as fully spin-polarized. However, one 
has to keep in mind that SOI entangles the spin and orbital momentum degrees of freedom 
thus reducing spin polarization. This is especially so for 
the currently investigated bulk TIs in which SOI is of electron-volt magnitude due 
to the presence of bismuth, the element with the largest atomic number $Z$ which has 
stable isotopes \cite{Carrier04,Wittel74}. 

In this Letter, we apply first-principles methodology to study the topological
surface states in Bi$_2$Se$_3$ and Bi$_2$Te$_3$, the ``second generation'' materials
which are currently considered as reference TIs \cite{Xia09,Zhang09}. This approach 
is free of empirical parameters and treats topological surface states on an equal 
footing with other states across the electronic spectrum. The degree of spin polarization 
of the topological surface states is found to be significantly reduced in both bismuth 
materials, and this reduction will affect most spintronic applications of these materials. 
We use our results to propose a way of controlling the degree of spin polarization 
associated with the charge current in thin TI slabs by applying a transverse electric
field. The proposed spintronic device is compared to a related proposal based on 
spin-polarized edge states in graphene \cite{Son06} and to conventional semiconductor 
quantum wells.

\begin{figure}[b]
\includegraphics[width=8cm]{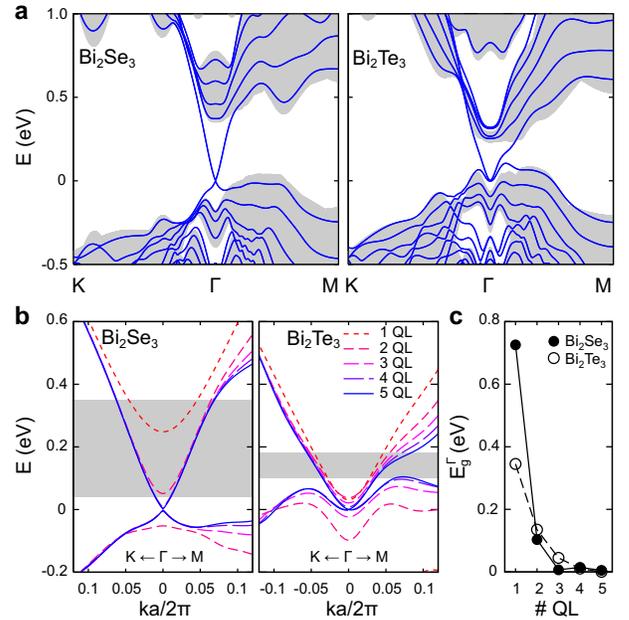}
\caption{\label{fig1}
(Color online). (a) Band structure of the 5 QL slabs of Bi$_2$Se$_3$ and Bi$_2$Te$_3$ (lines) 
superimposed with the projected band structure of the corresponding bulk materials
(shaded areas). (b) Evolution of the surface state band dispersion in the vicinity of $\Gamma$ 
point as a function of slab thickness. Shaded areas show the bulk band gaps. (c) Band gaps at 
the $\Gamma$ point induced by the interaction between the surface states as a function of slab thickness.
}
\end{figure}


First-principles electronic structure calculations have been performed within 
the density functional theory (DFT) framework employing the generalized gradient
approximation (GGA) to the exchange-correlation functional \cite{Perdew96}. Spin-orbit 
effects were treated self-consistently using fully relativistic norm-conserving 
pseudopotentials \cite{DalCorso05} acting on valence electron wavefunctions 
represented in the two-component spinor form \cite{Nordstrom96}. A plane-wave 
kinetic energy cutoff of 35~Ry has been employed for the wavefunctions. Surface states 
in Bi$_2$Se$_3$ and Bi$_2$Te$_3$ bulk TIs were investigated using (111) 
slab models of varying thickness (1--5 quintuple layers (QL) equivalent to $\sim$1--5~nm) 
with atomic positions taken from experimental data using a supercell geometry \cite{Cohen75}. 
Such an approach has already been used previously \cite{Xia09,Liu10,Eremeev10}.
The effects of applied electric field were simulated by adding a sawtooth-like potential profile with a constant slope 
in the slab region \cite{Kunc83}. The \textsc{pwscf} code of the \textsc{quantum-espresso}
distribution \cite{QE} was used in the present study. 


We start our discussion by considering the band structure of the bulk materials and slabs 
of finite thickness. Figure~\ref{fig1}(a) shows the band structure of 5 QL slabs 
superimposed with the bulk band structure projected onto the surface Brillouin zone (BZ). 
The band structures of the slab and bulk systems are aligned by matching the potential 
in ``bulk-like'' region of the slab to the bulk potential.
Within the DFT-GGA approach the band gaps of Bi$_2$Se$_3$ and Bi$_2$Te$_3$ are 0.31~eV and 0.08~eV, respectively. 
The value for Bi$_2$Se$_3$ is in good agreement with another theoretical report (0.32~eV \cite{Larson02})
and experimental data (0.35~eV \cite{Black57}), a fortuitous result considering the well-known tendency 
of DFT to underestimate band gaps \cite{Hybertsen86}. The calculated band gap for Bi$_2$Te$_3$ is consistent
with another first-principles result \cite{Youn01} and is smaller 
than 0.165~eV measured experimentally \cite{Chen09}. 

The topological surface states are distinguished in the slab band structures as 
the ``Dirac cone'' feature at the $\Gamma$ point, especially clear in the case of Bi$_2$Se$_3$. 
The Dirac point energies $E^{\rm D}$
(below, $E^{\rm D} = 0$~eV is set for convenience) are situated 0.04~eV and 0.10~eV below 
the corresponding bulk valence band maxima (VBM) within DFT-GGA. 
In slab geometry, the surface state bands are doubly degenerate with corresponding surface 
states of the opposite spin helicities localized at the opposite surfaces. The topological 
surface bands in bulk TIs do not respect electron-hole symmetry, and the Fermi velocity 
$v_{\rm F}$ shows rather complex behavior. At the VBM energy the calculated 
$v_{\rm F} = 4.6 \times 10^5$~m/s in Bi$_2$Se$_3$ agrees with experimental 
results \cite{Xia09}. The magnitude of $v_{\rm F}$ increases with increasing energy and reaches 
$6.4 \times 10^5$~m/s at the bulk conduction band minimum (CBM) energy for wavevectors 
along the $\Gamma$--K direction. However, along the $\Gamma$--M direction there is a noticeable 
softening ($v_{\rm F} =3.8 \times 10^5$~m/s) due to the hexagonal warping effect. 
The surface-state band of Bi$_2$Te$_3$ shows stronger anisotropy in agreement with experimental 
observations of a strong hexagonal warping \cite{Chen09}. At $E = 0.1$~eV, 
$v_{\rm F} = 4.1 \times 10^5$~m/s ($3.3 \times 10^5$~m/s) along the $\Gamma$--K ($\Gamma$--M) 
direction. The variation of $v_{\rm F}$ along the $\Gamma$--K direction is weak, while along 
the $\Gamma$--M direction it reduces to $1.2 \times 10^5$~m/s at the CBM energy.

Figure~\ref{fig1}(b) illustrates the evolution of the surface band dispersion as 
a function of slab thickness. Vacuum potential alignment is used for the comparison of the slab 
systems. In thin slabs, the interactions between the states localized at the opposite surfaces 
opens a gap ($E^\Gamma_g$) at the Dirac point.
The magnitude of the gap decays rapidly with increasing slab thickness [Fig.~\ref{fig1}(c)], 
and the values are in quantitative agreement with other first-principles 
results \cite{Liu10}. The dispersion of the surface state 
bands is essentially converged by 3~QL and 4~QL in Bi$_2$Se$_3$ and Bi$_2$Te$_3$, 
respectively. On the other hand, 
in thin slabs, quantum confinement pulls apart the bulk valence and conduction band edges thus 
increasing the energy range in which topological surface states are separated from bulk-like 
states. 
   
\begin{figure}
\includegraphics[width=8cm]{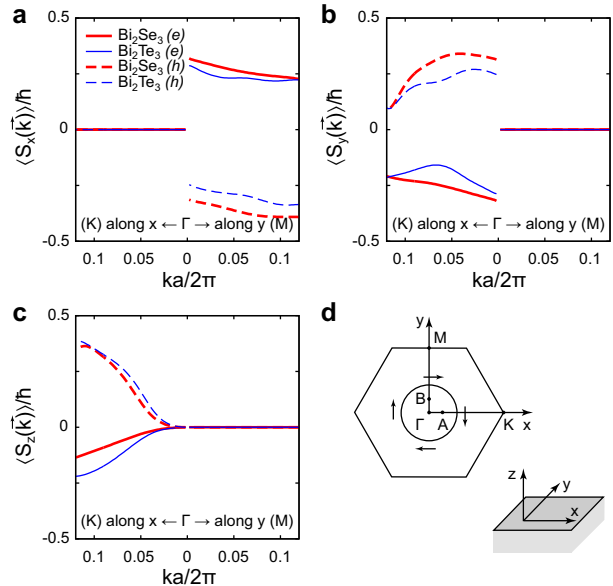}
\caption{\label{fig2}
(Color online). (a-c) Expectation values of the spin operators $\langle S_\alpha (\vec{k}) \rangle$ 
($\alpha = x,y,z$) for the topological surface states along the selected path ($A - \Gamma - B$) in 
momentum space (d) for 4~QL slabs of Bi$_2$Se$_3$ and Bi$_2$Te$_3$. Electron (solid lines) and 
hole (dashed lines) surface state bands are distinguished. The surface normal corresponds to $z$ direction.
Coordinate system is right-handed (d, inset).
}
\end{figure}


The striking feature of the surface states of TIs is the helicity of the spin polarization vector
$ \vec{P}(\vec{k}) = (2/\hbar) [ \langle S_x(\vec{k}) \rangle, \langle S_y(\vec{k}) \rangle, 
\langle S_z(\vec{k}) \rangle ] $ along the constant energy contours. Here, the expectation values of spin operators
$\langle S_{\alpha i}(\vec{k}) \rangle = (\hbar/2) \langle \psi_i (\vec{k}) | \sigma_\alpha | \psi_i (\vec{k}) \rangle$ ($\alpha = x,y,z$),
where $\psi_i (\vec{k})$ are the two-component spinor wavefunctions, $\sigma_\alpha$ the corresponding
Pauli matrices. The recent spin- and angle-resolved photoemission spectroscopy (spin-ARPES) experiments have
indeed indicated a one-to-one locking of the momentum and the direction of spin polarization vector 
pointing along $(\vec{k}\times\vec{z})$ \cite{Hsieh09}. First-principles results have confirmed this
picture \cite{Zhang10}. However, we would like to stress that the magnitude of spin-polarization $\vec{P}(\vec{k})$
can be reduced from the maximum value of 100\% since the electron spin quantum number is no longer
conserved in systems with SOI \cite{Kane05}.  Spin-orbit coupling leads to Bloch states that 
are not separately spin eigenstates but instead have entanglement between their spin and orbital 
parts.  An illustration of the effects of SOI on the spin projections in the case of bcc iron is 
given in Ref.~\onlinecite{Wang06}. 
Such effects are expected to be especially pronounced in bismuth materials where the 
exceptionally strong SOI, of the order of an electron-volt \cite{Carrier04,Wittel74}, 
would make the surface states a mixture of bulk states from a broad energy range. 

Figures~\ref{fig2}(a)--\ref{fig2}(c) show the calculated expectation values of the spin operators 
$\langle S_\alpha(\vec{k}) \rangle$ for both electron and hole states of the surface band 
in Bi$_2$Se$_3$ and Bi$_2$Te$_3$ along $x$ and $y$ directions in momentum space [see Fig.~\ref{fig2}(d) 
for definition]. The surface normal corresponds to $z$ direction and the standard convention of the 
right-handed coordinate system applies [Fig.~\ref{fig2}(d), inset]. We find that the spin polarization vectors 
are aligned preferentially in the $xy$ plane and the helicity is left (right) handed for the conduction or above $E^{\rm D}$
(valence or below $E^{\rm D}$) bands. This is consistent with previous experimental measurements \cite{Hsieh09} and theoretical
calculations \cite{Zhang10}. However, the magnitudes of the spin projections are significantly reduced
with respect to the nominal value of $\pm \hbar/2$. Close to the $\Gamma$ point we find $\sim$50--60\% 
spin-polarization for both materials. Its magnitude tends to decrease with increasing energy.
The magnitude of spin projections close to the Dirac point appears to be rather insensitive 
to the slab thickness (not shown here).  We have studied the microscopic origin of the reduction of spin polarization and find that the surface Bloch state is a complex superposition of different spin directions on different atoms; hence the reduction of the spin polarization cannot simply be understood in a tight-binding picture built from atomic eigenstates of total angular momentum.  A remarkable out-of-plane spin projection 
$\langle S_z(\vec{k}) \rangle$ develops in the $\Gamma$--K direction
as it was recently ascribed to the hexagonal warping effect \cite{Fu09}. In general, we find 
somewhat larger reduction of the spin polarization and more pronounced hexagonal warping effects on the 
out-of-plane spin projection in the case of heavier Bi$_2$Te$_3$.

\begin{figure}
\includegraphics[width=8cm]{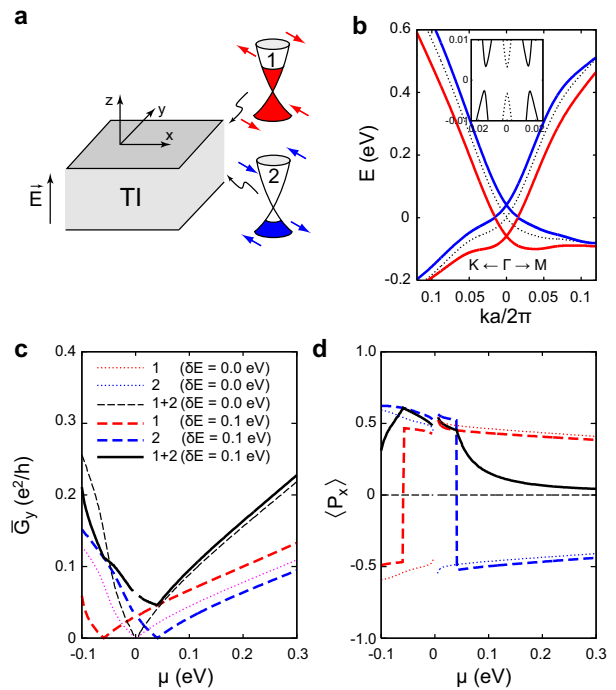}
\caption{\label{fig3}
(Color online). 
(a) Illustration of the effect of electric field $\vec{E}_{\rm ext}$ applied across a thin slab of bulk TI. 
The filling of the two surface state bands and the corresponding spin textures for electrons and holes
are shown. (b) Splitting of the surface state bands by applied electric field ($\delta E=0.1$~eV, 
solid lines) in a 3QL slab of Bi$_2$Se$_3$ compared to the band dispersion at zero field (dotted lines).
The inset shows a close-up view of band gaps at the charge neutrality point. (c) Individual (surfaces 1 and 2) and total ballistic 
conductances in $y$ direction per device width, $\overline{G}_y$, and (d) spin polarization along $x$ axis, $P_x$, 
for $\delta E=0.1$~eV and $\delta E=0$~eV as a function of chemical potential $\mu$ ($\mu=0$~eV corresponds to 
the charge neutrality point). The symbols have same meaning in (c) and (d).
}
\end{figure}


From the point of view of technological applications, an attractive feature of the TI materials 
is the intrinsic spin polarization of the current carried by the topological surface states. 
For charge carriers (electrons) moving along $y$ direction ($k_y > 0$) the transferred spin polarization 
(i.e. spin polarization of the current injected through a transparent tunneling barrier)
is along $x$ axis and its geometric average $\langle P_x (E) \rangle = (\pi/4)|\vec{P}(E)|$.
We define $\langle P_x (E) \rangle = \frac{ \int d\vec{k} v_y({\vec{k}}) P_x({\vec{k}}) \delta (E-E_{\vec{k}}) } { \int d\vec{k} v_y({\vec{k}}) \delta (E-E_{\vec{k}}) }$,
where $v_y({\vec{k}})$ is the carrier velocity along the transport direction. 
In the discussed materials $\langle P_x (E) \rangle \approx 50$\% and shows only weak 
dependence on $E$, but changes its sign when the character of charge carriers (electrons vs. holes) changes.
Moreover, in TI films the opposite spin helicities of the charge carriers at the opposite surfaces 
would result in zero net spin polarization.


We anticipate that improved control over spin transport can be achieved in few-nanometer thick
slabs of Bi$_2$Se$_3$ and Bi$_2$Te$_3$ which can be produced by molecular-beam epitaxy~\cite{Zhang09b,Li09}, vapor deposition 
\cite{Kong10,Kong10b}, or exfoliation \cite{Teweldebrhan10}. Tuning the chemical potential $\mu$ in 
such thin slabs by gating has already been achieved \cite{Kong10b}. Similarly to bilayer graphene 
\cite{Zhang09c}, the TI slabs are characterized by imperfect screening due to the semimetallic nature 
of the surface states. Thus, in the dual-gate device configuration or by coupling TI thin films to 
a ferroelectric substrate it is also possible to control splitting $\delta E = E_2^{\rm D} - E_1^{\rm D}$ 
between the Dirac point energies $E_1^{\rm D}$ and $E_2^{\rm D}$ of the two surface bands 
given with respect to the common chemical potential [Fig.~\ref{fig3}(a)]. Such splitting of  
$\sim$0.1~eV produced by the SiC substrate has been reported for Bi$_2$Se$_3$ thin films \cite{Zhang09b}.
Close to the Dirac point, 
the response of a thin TI film of thickness $d$ to an applied electric field $\vec{E}_{\rm ext}$ is expected to be dominated 
by the bulk screening contribution, $\delta E = |\vec{E}_{\rm ext}| d / \epsilon_{||}(0)$. The static dielectric constant 
along $c$-axis $\epsilon_{||}(0) = 75$ for Bi$_2$Te$_3$; probably smaller for Bi$_2$Se$_3$ \cite{LB}.
Figure~\ref{fig3}(b) shows the surface state band dispersion calculated from first-principles 
for a 3~QL slab of Bi$_2$Se$_3$ under applied electric field $\vec{E}_{\rm ext}$ resulting in $\delta E=0.1$~eV
(here, $\vec{E}_{\rm ext} = 1.72$~V/nm for the supercell dimension along $z$ axis equal to 4.36~nm).
The conical surface state bands shift rigidly with respect to each other while the bulk-like 
states (not shown) remain practically unaffected by the applied field. Figure~\ref{fig3}(c) shows 
the individual surface state Sharvin conductances along $y$ direction (i.e. the number of ballistic 
channels per device width along $x$ axis given in the units of lattice constant $a$), $\overline{G}_y^1$ and $\overline{G}_y^2$, 
and their sum $\overline{G}_y$.

The applied field lifts the semi-metallic conductance close to the charge 
neutrality point. Note that for both $\delta E=0.0$~eV  and $\delta E=0.1$~eV there is a 0.006~eV 
band gap opening at the charge-neutrality point due to the hybridization of the surface states localized on 
the opposite sides of the thin film. For $E_1^{\rm D} < \mu < E_2^{\rm D}$ the charge carriers are electrons (holes) 
at surface 1 (2). Importantly, in this range of $\mu$ the transferred spin polarization along $x$ direction 
is of the same sign for both surface conduction channels [Fig.~\ref{fig3}(d)]. The total transferred
spin polarization $\langle P_x \rangle$ is non-zero in the case of applied electric field. We note that these conditions are 
away from the symmetric situation that has been argued to lead to exciton condensation~\cite{Seradjeh08}.
The magnitude of $\langle P_x \rangle$ can be tuned by changing $\mu$ and achieves maximum its value $\langle P_x \rangle=0.608$ at 
$\mu = E_1^{\rm D}$. The sign of $\langle P_x \rangle$ can be changed by simply inverting the direction of $\vec{E}_{\rm ext}$.
That is, the dual-gate device based on a thin slab of TI permits to control independently both 
the conductance and the degree transferred spin polarization of injected current in a broad range, certainly 
a highly attractive opportunity for spintronics applications.


Such spintronic device can be compared to another theoretical proposal which realizes electric 
control over spin transport. This device based on zigzag graphene nanoribbons relies on the peculiar
spin-polarized edge states of topologically-trivial origin \cite{Son06}. While in TI slabs the degree 
of transmitted spin-polarization is limited by the SOI and by the geometric average prefactor $\pi/4$,
the localized states at the zigzag edges of graphene are always 100\% spin-polarized. On the 
other hand, the degree of spin-polarization can be continuously tuned in the TI slabs. The direction
of $\vec{P}$ is fixed with respect to the current direction at the TI surfaces, while graphene-based 
magnetic systems are characterized by low magnetic anisotropy as a result of very weak SOI 
in carbon and by relatively short spin correlation length \cite{Yazyev08,Yazyev10}.

In a more general context, it is instructive to compare the spin transport effects at the TI surfaces 
to those in other 2D systems, e.g. to the corresponding effect induced by Rashba SOI in a 2D electron gas, where a current also induces a spin density.
The basic difference is that the Rashba coupling leads to two Fermi surfaces with opposite spin directions at each momentum, and the resulting current-induced spin densities nearly cancel if the SOI is small.  
A single TI surface has only {\it one} spin state at each momentum and hence no cancellation. Quantitatively, 
we write the Rashba SOI as $\alpha \hbar ({\vec k} \times \sigma)_z$ where $\alpha$ has units of velocity 
and ${\bf \sigma}$ are Pauli matrices.  The cancellation between Fermi surfaces reduces the spin density 
by a factor of order $\alpha / v_{\rm F}$ compared to a model TI surface with a single Dirac cone, where 
$v_{\rm F}$ is the Fermi velocity.  Taking one example, $\alpha$ in a InGaAs/InAlAs quantum wells was measured 
to be $7$--$14 \times 10^3$ m/s \cite{Nitta97}, more than an order of magnitude less than 
$v_{\rm F} \sim 5 \times 10^5$~m/s in Bi$_2$Se$_3$ and Bi$_2$Te$_3$ \cite{Xia09,Chen09}.


We would like to thank E. Kioupakis, J. Orenstein and C. Jozwiak for discussions. 
This work was supported in part by National Science Foundation Grants No.~DMR07-05941 and DMR08-04413.
O.\ V.\ Y. is recipient of a Swiss National Science Foundation fellowship (grant No.~PBELP2-123086) and
partially supported by the Director, Office of Science, Office of Basic Energy Sciences, 
Division of Materials Sciences and Engineering Division, U.S. Department 
of Energy under Contract No.~DE-AC02-05CH11231.  
Computational resources have been provided by TeraGrid (Kraken).

\end{document}